\documentclass[amsfonts,amsmath,amssymb,showpacs,floatfix,superscriptaddress,nofootinbib,preprintnumbers,11pt]{article}
\usepackage{amsmath,amsthm,amssymb}
\usepackage{graphicx}
\begin{document}
\title{\textbf{\textsf{A single model of interacting dark energy:
generalized phantom energy or generalized Chaplygin gas}}}
\author{ \textsf{Mubasher Jamil\footnote{mjamil@camp.nust.edu.pk}}
 \\ \\
\small Center for Advanced Mathematics and Physics, Campus of
College of E\&ME,\\ \small National University of Sciences and
Technology, \small Peshawar Road, Rawalpindi - 46000, Pakistan
 }\maketitle
\begin{abstract}
I present a model in which dark energy interacts with matter. The
former is represented by a variable equation of state. It is shown
that the phantom crossing takes place at zero redshift, moreover,
stable scaling solution of the Friedmann equations is obtained. I
show that dark energy is most probably be either generalized phantom
energy or the generalized Chaplygin gas, while phantom energy is
ruled out as a dark energy candidate.
\end{abstract}

 \large
\newpage
\section{Introduction}

Numerous cosmological observations show that our universe is
pervaded with a mysterious dark energy which constitutes more then
seventy percent of the cosmic energy density \cite{perl}. The
presence of dark energy is essential to get a spatially flat
universe as is observed today. Although dark energy is dominant in
the universe, however we don't know its composition and origin.
Theoretically there are several candidates that can account for dark
energy including the cosmological constant, quintessence, Chaplygin
gas and phantom energy etc, to name a few. However no single
candidate among these can completely and satisfactorily meet the
observational data. Another problem associated with it is that the
exact form of the equation of state (EoS) of dark energy is yet to
be discovered. In literature, several exotic EoS are proposed (see
for example \cite{barb}). We in this paper will discuss the dynamics
of dark energy with a recently proposed EoS that combines three
forms of dark energy equations of state including generalized
Chaplygin gas, phantom energy and its generalized form. Thus this
single model presents unified cosmological model of analyzing dark
energy dynamics.

From the cosmological observations, it is now clear that there are
three major cosmic ingredients including dark matter, radiation and
dark energy. It is naturally expected that these cosmic species may
entertain some form of interaction at some stage in the evolution of
the universe. This interaction is specifically important in the
present cosmic setting when we observe that the energy densities of
dark matter and dark energy are almost of the same orders of
magnitude. Thus it excludes radiation from the interacting system.
Recently there have been several cosmological models build up having
dark energy of the forms mentioned above interacting with matter
\cite{peebles}. It is interesting to note that most of these models
predict matter dominated scenario in the late evolution of the
universe where dark energy decays into matter. This conclusion
apparently avoids the imminent cosmic singularities like the big rip
or the big crunch. Thus it naturally predicts a cyclic form of the
evolution of the universe from matter domination to dark energy
domination, back and forth.

In the present paper, I model my system by first assuming the
spatially flat background spacetime. Further, I assume the spacetime
to contain only dark matter and dark energy and having exotic
interaction. Both dark matter and dark energy are specified by their
respective equations of state. I choose a specific form of the
interaction term that is well motivated from the dimensional
considerations. The exact form of the interaction term will be
originally motivated from the theory of quantum gravity which is as
yet in the process of construction by a number of theorists. That
term contains a coupling parameter whose numerical value determines
the strength of the interaction while its signature describes the
direction of flow of the energy. I shall assume its numerical value
to be between 0 and 1 while its signature to be positive (showing
flow of energy from dark energy to matter). Finally I perform
stability analysis of the dynamical system and discuss my results.

\section{Modeling of dynamical system}
I start by assuming the background to be spatially flat, homogeneous
and isotropic Friedmann-Robertson-Walker (FRW) spacetime
\begin{equation}
ds^2=-dt^2+a^2(t)[dr^2+ r^2(d\theta^2+\sin^2\theta d\phi^2)].
\end{equation}
The equations of motion corresponding to FRW spacetime filled with
the two component fluid (matter and dark energy) are
\begin{eqnarray}
\dot{H}&=&-\frac{\kappa ^{2}}{6}(p_{de}+\rho _{de}+\rho _{m}),\\
H^{2}&=&\frac{\kappa ^{2}}{3}(\rho _{de}+\rho _{m}).
\end{eqnarray}
Here $\kappa ^{2}=8\pi G$ is the Einstein's gravitational constant
and $H=H(t)\equiv\dot a/a$ is the Hubble parameter. Also subscripts
`m' and `de' correspond to matter and dark energy respectively while
$p$ and $\rho$ are respectively pressure and energy density. In Eq.
(2), matter is assumed to be pressureless ($p_m=0$) while the dark
energy's equation of state is considered of the form \cite{peter}
\begin{equation}
p_{de}=-Br^{2(\alpha-1)}\rho_{de}^\alpha, \ \ \alpha\neq0,
\end{equation}
where $B$ is a positive constant and $\alpha$ can be either positive
or negative. Note that $\alpha=1$ represents phantom energy,
$\alpha>0$ gives generalized phantom energy while $\alpha<0$
represents generalized Chaplygin gas. Thus the generalized phantom
energy and Chaplygin gas models can be unified by the above equation
of state.

The energy conservation equation for the dynamical system under
consideration is
\begin{equation}
\dot\rho_{de}+\dot\rho_m+3H(\rho_m+\rho_{de}+p_{de})=0.
\end{equation}
Due to interaction between the two components, the energy
conservation would not hold for the individual components, therefore
the above conservation equation will break into two equations:
\begin{eqnarray}
\dot{\rho }_{de}+3H(p_{de}+\rho _{de})&=&-Q,\\
\dot{\rho }_{m}+3H\rho _{m}&=&Q.
\end{eqnarray}
Here $Q$ is the interaction term which in general is a function of
energy densities and the Hubble parameter i.e.
$Q(H\rho_m,H\rho_{de})$ which upon Taylor expansion yields the form
given below in (8). This interaction is demanded to be positive to
avoid the violation of the laws of thermodynamics \cite{campo}. I
also insert a dimensionless coupling parameter $c$ in $Q$ to
determine the strength of the interaction, thus I have
\begin{equation}
Q=3Hc(\rho _{de}+\rho _{m}).
\end{equation}
To study the dynamics of the system, I proceed by setting
\begin{equation}
x=\ln a=-\ln (1+z),
\end{equation}
which is termed as the e-folding time parameter and $z$ is the
redshift parameter. Moreover, the density and pressure of dark
energy can be expressed by dimensionless variables $u$ and $v$ as
\begin{equation}
u=\Omega _{de}=\frac{\rho _{de}}{\rho _{cr}}=\frac{\kappa ^{2}\rho
_{de}}{ 3H^{2}},\ \ v=\frac{\kappa ^{2}p_{de}}{3H^{2}}.
\end{equation}
The EoS parameter $\omega_{de} $ is conventionally defined by
\begin{equation}
\omega_{de}\equiv \frac{p_{de}}{\rho _{de}},
\end{equation}
which after using (10) becomes
\begin{equation}
\omega_{de}=\frac{v}{u}.
\end{equation}
The density parameters of dark energy and dark matter are related as
\begin{equation}
\Omega _{m}=\frac{\kappa ^{2}\rho _{m}}{3H^{2}}=1-\Omega _{de}=1-u.
\end{equation}
Using Eqs. (8)-(13) in (6) and (7), I obtain the following
autonomous system
\begin{eqnarray}
\frac{du}{dx}&=&-3c-3v+3uv,\\
\frac{dv}{dx}&=&-3\alpha
v\Big(1+\frac{v}{u}+\frac{c}{u}\Big)+3v(v+1).
\end{eqnarray}
Note that the above dynamical system matches with \cite{wu} if
$\alpha<0$. Their it is shown that interacting GCG can explain late
time acceleration. I am here particularly interested in the cases
when $\alpha=1$ and $\alpha>0$. The stability of this system is
performed by first finding its critical points i.e. the points for
which the left hand sides of the system vanish identically. Thus I
obtain the only critical point:
\begin{eqnarray}
u_c&=&1-c,\\
v_c&=&-1.
\end{eqnarray}
It needs to be checked whether the system is stable about
($u_c,v_c$). For this purpose, I linearize the system by
substituting $u=u_c+\delta u$, and $v=v_c+\delta v$, where $\delta
u$ and $\delta v$ are small perturbations about the critical point.
I finally obtain
\begin{eqnarray}
\frac{d\delta u}{dx}&=&3v_c\delta u+3(-1+u_c)\delta v,\\
\frac{d\delta v}{dx}&=&\frac{3}{u_c^2}[\alpha(v_c^2+cv_c)\delta
u+[(1+2v_c-\alpha)u_c^2-2\alpha u_c v_c-\alpha cu_c]\delta v].
\end{eqnarray}
Now the given dynamical system is stable about the critical point if
the real parts of the eigenvalues are both negative. In this case,
the critical point is called a stable node. However, if one is
positive then such a point is termed a saddle point. If both
positive, it is called an unstable node. The eigenvalues are
\begin{eqnarray}
\lambda_1&=&-3-\frac{-3\alpha+3\sqrt{\alpha(\alpha-4c(1-c))}}{2(1-c)},\\
\lambda_2&=&-3-\frac{-3\alpha-3\sqrt{\alpha(\alpha-4c(1-c))}}{2(1-c)}.
\end{eqnarray}
It can be shown easily that the real parts of both the eigenvalues
are negative. In Figures 1 and 2, I show phase space diagrams of the
solutions of my dynamical system for different choices of initial
conditions. Here I took $\alpha=0.07$ and $\alpha=1$, representing
generalized phantom energy and the phantom energy, respectively.
Note that I am not discussing the case of $\alpha<0$ as it already
has been discussed in \cite{wu}. In figures 3 and 4, I provide
pictorial representation of the solutions against the e-folding time
parameter. It is interesting to note that the solutions are stable
(without fluctuations or oscillations of any kind). Figure 5 shows
an unstable solution which is consequently of not much physical
interest. However it puts an upper limit on the choice of $c=0.6$ if
the dark energy is of phantom type $\alpha=1$. In Figures 6 and 7, I
plot the dark energy state parameter against e-folding time
parameter $x$. The former one presents a smooth transition from the
quintessence to phantom regime while this transition takes place at
$x=0$, which corresponds to the present epoch. The later figure
shows a rather stale case where the state parameter remains
unchanged throughout the variation of $x$. I also point out here
that this last result is unaffected for various values of $c$. Hence
it can be said that dark energy can most probably be either
generalized Chaplygin gas or the generalized phantom energy.

\newpage
\section{Conclusion}

Models of interacting dark energy possess tremendous potential to
alleviate several cosmological puzzles, most notably the cosmic
coincidence problem and the phantom crossing scenario. The model
predicts a dynamical nature of dark energy where its equation of
state and energy density both vary over cosmic time which turns out
to be consistent with the observations. As seen in the previous
section, the phantom crossing scenario is well explained if the
coupling parameter is non-zero but very small positive number
compared to unity. In this paper, I have analyzed the model of
interacting dark energy using an interesting equation of state which
was originally specified in the context of wormhole physics, very
similar to the Chaplygin gas that was originally motivated from an
engineering problem. The equation of state amalgamates three
different forms of dark energy and therefore its inclusion in the
interacting model makes the model more comprehensive and unified.
From the analysis of this paper, it can be concluded that dark
energy could most probably of the form of generalized phantom energy
or generalized Chaplygin gas while phantom energy is ruled out.
Also, this work serves as a generalization of earlier studies in
this direction \cite{zhang,wu,jamil} which dealt with Chaplygin gas
only.

\newpage
\begin{figure}
\includegraphics[scale=.7]{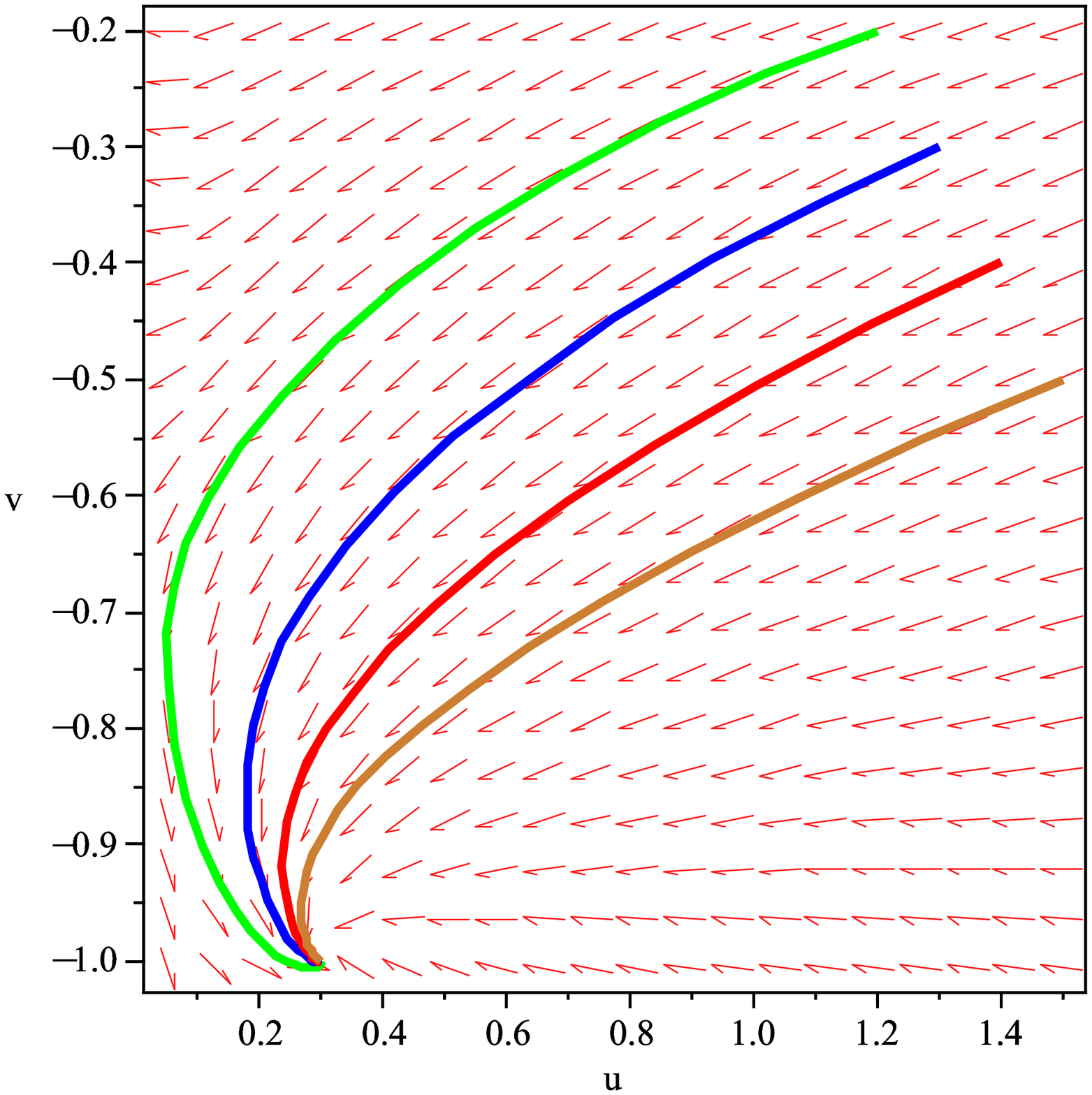}\\
\caption{The phase diagram of the interacting dark energy model with
$c=0.7$ and $\alpha=0.07$. The curves correspond to the initial
conditions $u(-2)=1.2, v(-2)=-0.2$ (green); $u(-2)=1.3, v(-2)=-0.3$
(blue); $u(-2)=1.4, v(-2)=-0.4$ (red); $u(-2)=1.5, v(-2)=-0.5$
(brown).}
\end{figure}
\newpage
\begin{figure}
\includegraphics[scale=.7]{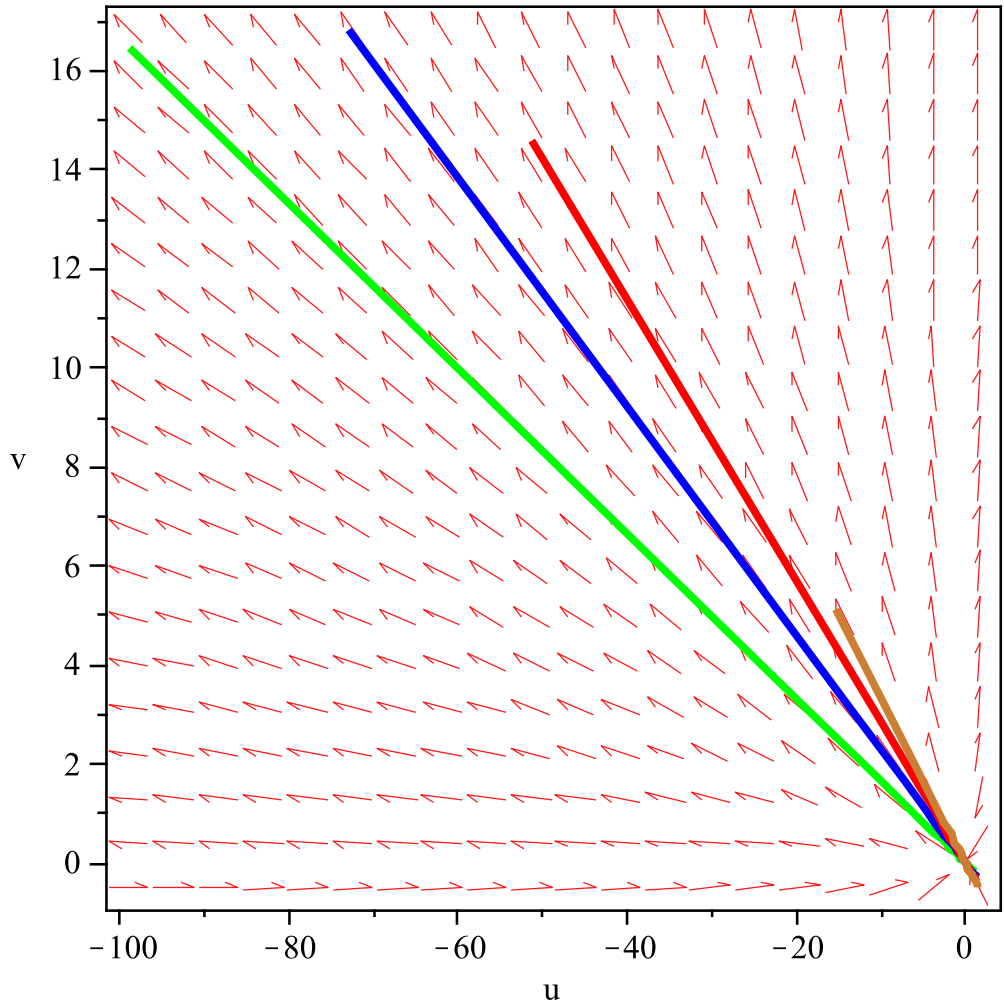}\\
\caption{The phase diagram of the interacting dark energy model
model with $c=0.8$ and $\alpha=1$. The curves correspond to the
initial conditions as given in Fig. 1.}
\end{figure}
\newpage
\begin{figure}
\includegraphics[scale=.7]{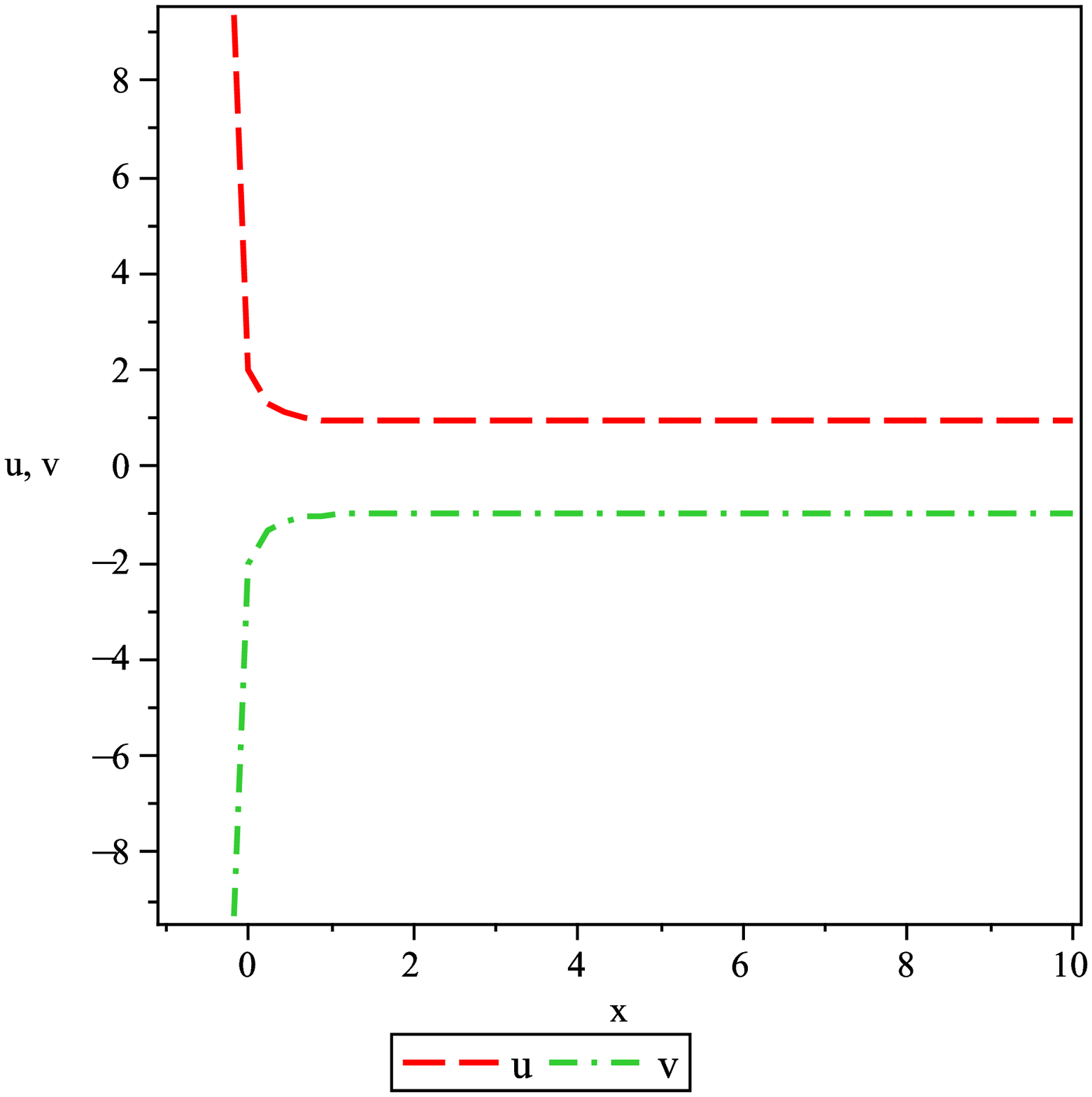}\\
\caption{The parametric functions $u$ and $v$ are plotted against
the e-folding time parameter $x$. The parameters are fixed as
$c=0.7$ and $\alpha=0.07$. The initial condition is $u(0)=2,
v(0)=-2$. }
\end{figure}
\newpage
\begin{figure}
\includegraphics[scale=.7]{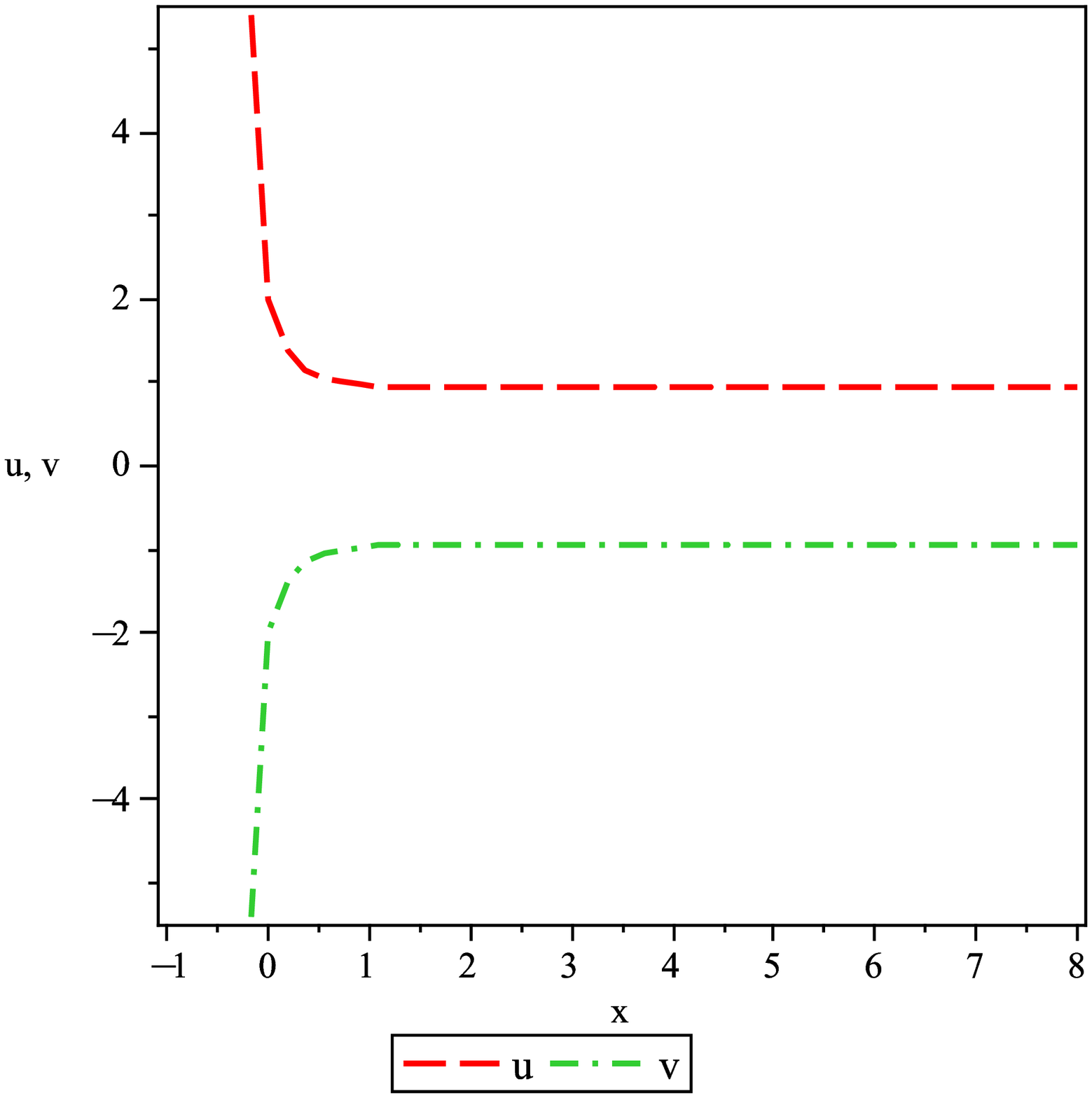}\\
\caption{The parametric functions $u$ and $v$ are plotted against
the e-folding time parameter $x$. The parameters are fixed as
$c=0.06$, $\alpha=1$. The initial condition is $u(0)=2, v(0)=-2$.}
\end{figure}
\newpage
\begin{figure}
\includegraphics[scale=.7]{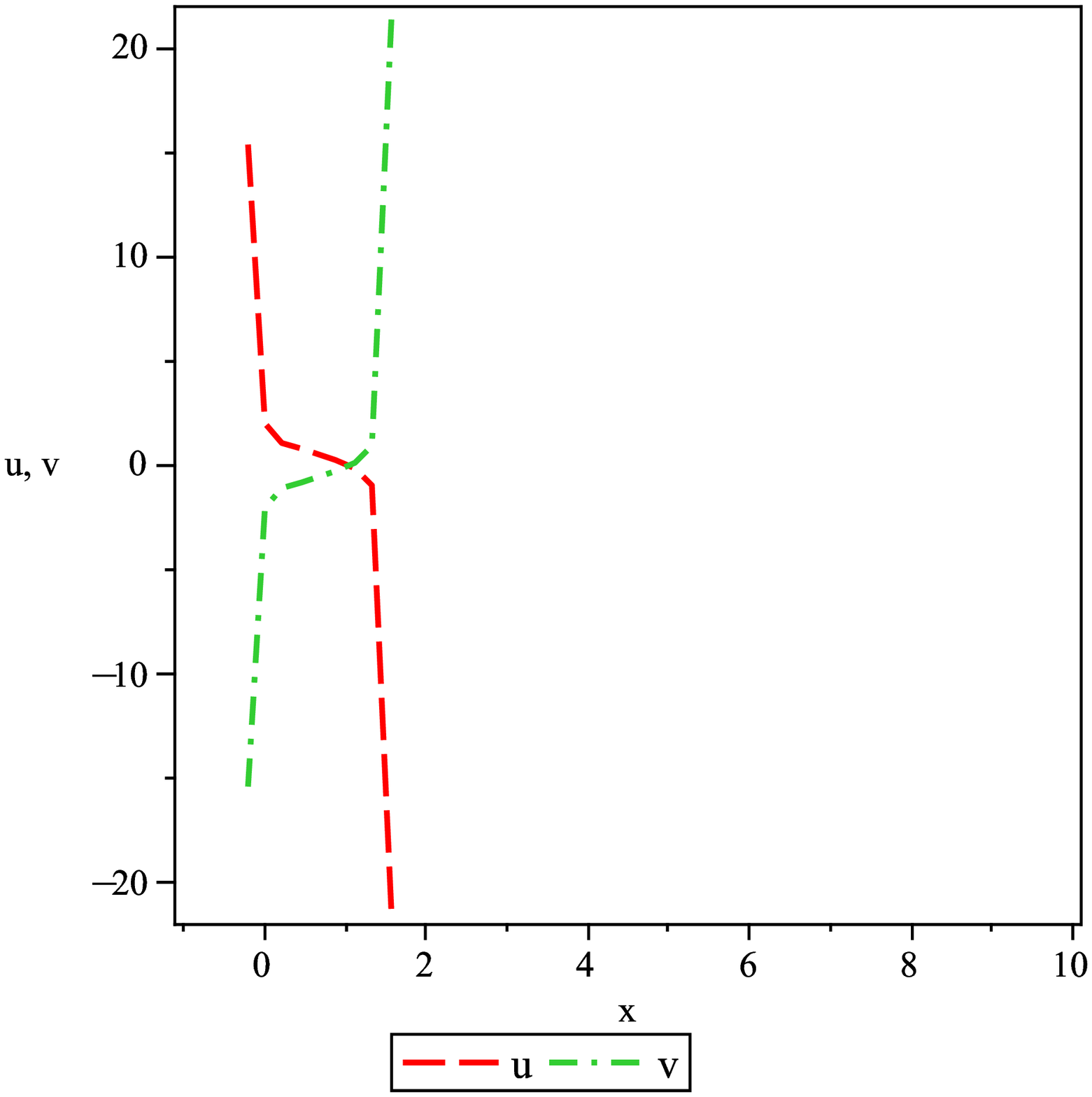}\\
\caption{The parametric functions $u$ and $v$ are plotted against
the e-folding time parameter $x$. The parameters are fixed as
$c=0.6$, $\alpha=1$. The initial condition is $u(0)=2, v(0)=-2$.}
\end{figure}
\newpage
\begin{figure}
\includegraphics[scale=.7]{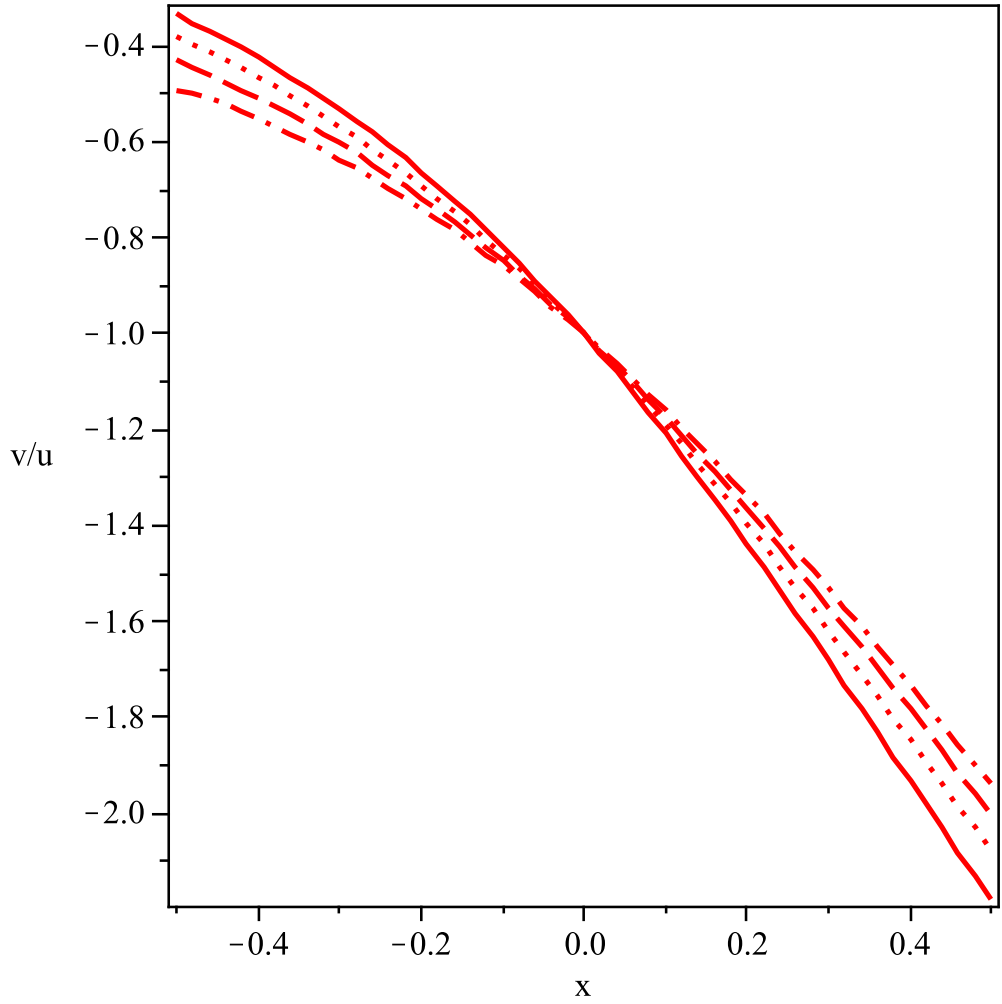}\\
\caption{The EoS parameter $\omega_X=v/u$ is plotted against $x$.
The parameters are fixed as $c=0.7$ and $\alpha=0.07$. The initial
conditions are $u(0)=1, v(0)=-1$ (solid line); $u(0)=1.1,v(0)=-1.1$
(dots); $u(0)=1.2,v(0)=-1.2$ (dashes); $u(0)=1.3,v(0)=-1.3$ (dash
dot).}
\end{figure}
\newpage
\begin{figure}
\includegraphics[scale=.7]{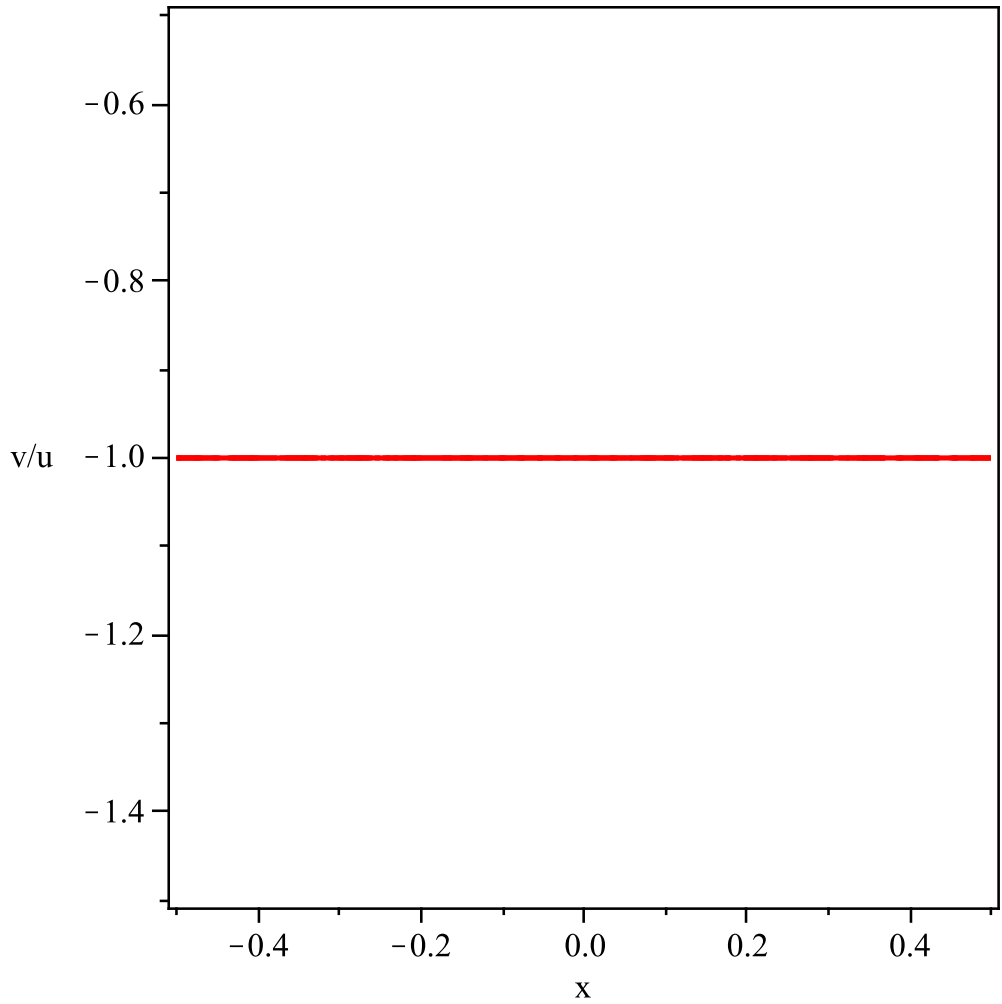}\\
\caption{The EoS parameter $\omega_X=v/u$ is plotted against $x$.
The parameters are fixed as $c=0.5$ and $\alpha=1$. The initial
conditions are the same as in Fig. 6}
\end{figure}
\end{document}